\documentclass[12pt]{iopart}
\usepackage{graphicx}
\usepackage{color}

\newcommand{\rmint}{\mathrm{int}}

\newcommand{\calN}{\mathcal{N}}
\def\bvarrho{\mbox{\boldmath{$\varrho$}}}
\def\bepsilon{\mbox{\boldmath{$\epsilon$}}}
\newcommand{\rmor}{\mathrm{or}}

\newcommand{\rmtot}{\mathrm{tot}}

\newcommand{\rmg}{\mathrm{g}}

\newcommand{\rmwith}{\mathrm{with}}
\newcommand{\rmand}{\mathrm{and}}

\begin{document}
%\begin{center}
\title[Time-dependent pointer states of the quantized atom-field model in a nonresonance regime]{Time-dependent pointer states of the quantized atom-field model in a nonresonance regime and consequences regarding the decoherence of the central system}
%\end{center}
\author{Hoofar Daneshvar and G W F Drake}
\address{Department of Physics, University of Windsor, Windsor ON, N9B 3P4, Canada}
\ead{hoofar@uwindsor.ca and GDrake@uwindsor.ca}
\begin{abstract}
We consider the quantized atom-field model and for the regime that $\hat{H}_{\cal E}\ll\hat{H}_{\cal S}\ll\hat{H'}$ (but $\hat{H}_{\cal E}\neq0$ and $\hat{H}_{\cal S}\neq0$); where $\hat{H}_{\cal E}$, $\hat{H}_{\cal S}$ and $\hat{H'}$ respectively represent the self Hamiltonians of the environment and the system, and the interaction between the system and the environment. Considering a single-mode quantized field we obtain the time-evolution operator for the model. Using our time-evolution operator we calculate the time-dependent pointer states of the system and the environment (which are characterized by their ability not to entangle with states of another subsystem) by assuming an initial state of the environment in the form of a Gaussian package in position space. We obtain a closed form for the offdiagonal element of the reduced density matrix of the system and study the decoherence of the central system in our model. We will show that for the case that the system initially is not prepared in one of its pointer states, the offdiagonal element of the reduced density matrix of the system will decay with a decoherence time which is inversely proportional to the square root of the mass of the bosonic field particles.
\end{abstract}

\maketitle

\section{Introduction}
\subsection{Foreword}
In our previous paper \cite{paper1} we discussed the pointer states of measurement\footnote{The pointer states of a subsystem are characterized by their ability not to entangle with the states of another subsystem and appear in the diagonal state of the total composite system after premeasurement by the environment. As we elaborately described in \cite{paper1}, generally we should
distinguish between the pointer states of a system and the preferred basis of measurement. We proved that the pointer states of a subsystem generally are time-dependent and a preferred basis of measurement does not exist, unless under some specific conditions (discussed there in \cite{paper1}) for which the pointer states of measurement become time-independent. Moreover, the pointer states of a system necessarily are not orthonormal amongst themselves at all times. Therefore, necessarily they cannot form a basis for the Hilbert space of the system at all times.} and we presented a general method for obtaining the pointer states of a two-level system and its environment, for a given total-Hamiltonian defining the system-environment model\footnote{Although a reference to our other works (\cite{paper1} and \cite{paper3}) can be useful for the interested reader, in writing this paper we have tried to make it self-contained; so that the reader can well understand this work without a need to refer to the other two works.}. As we discussed in this paper, time-independence of pointer states by no means should be taken for granted; since time-independent pointer states can be realized only under some specific conditions \cite{paper1}. We used our method in order to rederive the time-dependent pointer states of the system and the environment (initially prepared in the coherent state) in the Jaynes-Cummings model (JCM) of quantum optics and for the exact resonance regime; verifying the previous results for the JCM \cite{Gea-Banacloche,Gea-Banacloche2}. Also, to further demonstrate the generality and usefulness of our method of obtaining pointer states, in another paper \cite{paper3} we obtained the time-dependent pointer states of the system and the environment for the generalized spin-boson model and in the exact resonance regime.

In this paper we study the quantized atom-field model \emph{without} the assumption of resonance between the splitting of the states of the two-level atom $\omega_{0}$ and the cavity eigenmode frequency $\omega$. Our quantized atom-field model basically is consisted of a two-level atom, with upper and lower levels that can respectively be represented by $|a\rangle$ and $|b\rangle$, interacting with a single-mode quantized bosonic field (such as photons) inside an ideal cavity, represented by creation and annihilation operators $\hat{a}^{\dag}$ and $\hat{a}$. The Hamiltonian for the total composite system can be written as \cite{Scully}
\begin{equation}\label{1}
\hat{H}=\frac{1}{2}\omega_{0}\hat{\sigma}_{z}+\omega\hat{a}^{\dag}\hat{a}+\rmg \chi \hat{\sigma}_{x} \hat{x},
\end{equation}
Where $\rmg=-\bvarrho_{12}.\bepsilon\sqrt{\frac{\omega_{0}}{2\hbar\varepsilon_{\circ}V}}$ is the atom-field coupling constant, with $\bvarrho_{12}=e\langle a|\textbf{r}|b\rangle $ as the atomic electric-dipole transition matrix element. ($\bepsilon$ is the field polarization vector and V is the cavity
mode volume). Also
$\chi=\sqrt{2m\omega}$; so that $\chi\hat{x}=\hat{a}+\hat{a}^{\dag}$.\footnote{Here in this paper we use the atomic units wherein $\hbar=1$.}

The main purpose of this paper is to obtain the time-dependent pointer states of the system and the environment, as well as expressions for the evolution of the reduced density matrix of the system in the regime that $\hat{H}_{\cal E}\ll\hat{H}_{\cal S}\ll\hat{H'}$, but $\hat{H}_{\cal S}\neq0$ and $\hat{H}_{\cal E}\neq0$. In other words, to demonstrate how our formulation
for obtaining time-dependent pointer states can be used in practice, here we consider a very specific regime of the parameter space and will obtain the corresponding pointer states of the system and the environment within that specific regime; as pointer states (\emph{if} they exist in certain regimes of a system-environment model) generally depend on the specific regime of the parameter space which we are considering and generally acquire different forms in different regimes of the parameter space, \emph{even} for a specifically given system-environment Hamiltonian.

For the Hamiltonian of equation (\ref{1}), as we will show here, the special regime of $\hat{H}_{\cal E}\ll\hat{H}_{\cal S}\ll\hat{H'}$ is valid only and only if we have
\begin{equation}\label{1.5}
1\ll\sqrt{\frac{\omega_{0}}{\omega}}\ll|\bvarrho_{12}.\bepsilon|\times\sqrt{\frac{m}{\hbar\varepsilon_{\circ}V}}\ , \quad \omega\neq0 \quad \rmand \quad \omega_{0}\neq0.
\end{equation}
To show this, note that the condition $\hat{H}_{\cal S}\ll\hat{H'}$ requires that $\sqrt{\frac{\omega_{0}}{\omega}}\ll|\bvarrho_{12}.\bepsilon|\times\sqrt{\frac{m}{\hbar\varepsilon_{\circ}V}}$; while the condition $\hat{H}_{\cal E}\ll\hat{H}_{\cal S}$ requires that $1\ll\sqrt{\frac{\omega_{0}}{\omega}}$. Also, we should emphasize that we \emph{must} have $\omega\neq0$ ($\hat{H}_{\cal E}\neq0$) and $\omega_{0}\neq0$; as otherwise we would have zero coupling $\rmg\chi$, and we cannot have $\hat{H}_{\cal S}\ll\hat{H'}$ (since we have $\rmg\chi=-\bvarrho_{12}.\bepsilon\sqrt{\frac{m\omega\omega_{0}}{\hbar\varepsilon_{\circ}V}}$). Therefore, as we see, the regime that we are considering and the results of this article are valid \emph{only} in the specific part of the parameter space where the inequalities of equation (\ref{1.5}) are valid.

\noindent This paper is organized as follows:

After this foreword we review our method for obtaining the pointer states of the system and the environment; and in section 3 we exploit it in order to calculate the time-dependent pointer states of the quantized atom-field model represented by the Hamiltonian of equation (\ref{1}).

In order to be able to exploit our method and obtain the pointer states of the system and the environment in our model, we need to know the time-evolution operator of our model in the regime that we are considering. This task is done in section 2.

In section 4 we exploit the pointer states of the system and the environment (which we obtain in section 3) in order to study the decoherence of the central system in our model. Finally, in section 5 we further discuss the significance of our results and the conclusions.

\subsection{Review of the method}
In order to be able to obtain the pointer states of the system and the environment for an arbitrary total Hamiltonian defining our system-environment model we first need to find those probable initial states of the system which do not entangle with the states of the environment throughout their evolution with time; and then we should obtain their time evolution. Finally, we should obtain their corresponding states from the environment which in fact, are the pointer states of the environment. As we saw in \cite{paper1}, existence of pointer states may require having a sufficiently large environment which contains a large number of degrees of freedom. In other words, pointer states characterized by their ability not to entangle with the states of another subsystem, do not necessarily exist in every arbitrary regime.

Consider a two-state system $\cal S$ with two arbitrary basis states $|a\rangle$ and $|b\rangle$, initially prepared in the state
\begin{equation}\label{240}
|\psi^{\cal S}(t_{0})\rangle=\alpha|a\rangle+\beta|b\rangle \quad \rmwith \quad |\alpha|^{2}+|\beta|^{2}=1;
\end{equation}
and an environment initially prepared in the state
\begin{equation}\label{250}
|\Phi^{\cal E}(t_{0})\rangle=\sum_{n=0}^{\infty}c_{n}|\varphi_{n}\rangle,
\end{equation}
where $\{|\varphi_{n}\rangle\}$'s are a complete set of basis states for the environment. For the two-state system with the two basis states $|a\rangle$ and $|b\rangle$ we can take the set of any four linearly independent operators in the Hilbert space of the system as a complete set of basis operators, which can induce any change to the initial state of the two-state system given by equation (\ref{240}). For example, we can take the Pauli operators in addition to the identity operator $\hat{I}=|a\rangle\langle a|+|b\rangle\langle b|$ as our complete set of basis operators; or equivalently we can take the four operators $|a\rangle\langle a|$, $|a\rangle\langle b|$, $|b\rangle\langle a|$ and $|b\rangle\langle b|$ as our complete set of basis operators. So, the time evolution operator for the global state of the system and the environment, which (for a two-state system) generally is of the form
\begin{equation}\label{260}
\hat{U}_{\rm tot}(t)=\sum_{\alpha=1}^{4} \hat{S}_{\alpha}\otimes\hat{\cal E}_{\alpha}\ ,
\end{equation}
Can be considered as
\begin{equation}\label{270}
\hat{U}_{\rm tot}(t)=\hat{\cal E}_{1}|a\rangle\langle a|+\hat{\cal E}_{2}|a\rangle\langle b|
+\hat{\cal E}_{3}|b\rangle\langle a|+\hat{\cal E}_{4}|b\rangle\langle b|.
\end{equation}
In the above equation $\hat{\cal E}_{i}$'s are operators acting on the Hilbert space of the environment, and depend on the total Hamiltonian defining the system-environment model. For example, for the Jaynes-Cummings model and for exact resonance and in the rotating wave approximation (RWA), it can be shown \cite{Scully} that the $\hat{\cal E}_{i}$'s are given by the following relations
\begin{eqnarray}\label{280}
\nonumber \hat{\cal E}_{1}=\cos(\rmg t\sqrt{\hat{a}^{\dag}\hat{a}+1})\ , \quad \hat{\cal E}_{2}=-i\ \frac{\sin(\rmg t\sqrt{\hat{a}^{\dag}\hat{a}+1}}{\sqrt{\hat{a}^{\dag}\hat{a}+1}}\ \hat{a})\\
\hat{\cal E}_{3}=-i\hat{a}^{\dag}\ \frac{\sin(\rmg t\sqrt{\hat{a}^{\dag}\hat{a}+1}}{\sqrt{\hat{a}^{\dag}\hat{a}+1}})\ , \quad
\hat{\cal E}_{4}=\cos(\rmg t\sqrt{\hat{a}^{\dag}\hat{a}}).
\end{eqnarray}

Using equations (\ref{240}) to (\ref{270}) we can write the global state of the system and the environment as follows
\begin{eqnarray}\label{290}
\nonumber |\Psi^{\rm tot}(t)\rangle=\hat{U}_{\rm tot}(t).\ (\alpha|a\rangle+\beta|b\rangle)\otimes(\sum_{n=0}^{\infty}c_{n}|\varphi_{n}\rangle)\\
=\textbf{A}(t)\ |a\rangle+\textbf{B}(t)\ |b\rangle\; \quad \rmwith \quad
\textbf{A}(t)=\sum_{n=0}^{\infty}c_{n}\{\alpha\hat{\cal E}_{1}+\beta\hat{\cal E}_{2}\}\ |\varphi_{n}\rangle \\ \rmand \quad \nonumber
\textbf{B}(t)=\sum_{n=0}^{\infty}c_{n}\{\alpha\hat{\cal E}_{3}+\beta\hat{\cal E}_{4}\}\ |\varphi_{n}\rangle.
\end{eqnarray}
In order to find those probable initial states of the system which do not entangle with the states of the environment we first define $\hat{G}(t)$ as the operator in the Hilbert space of the environment which relates the vectors $\textbf{A}(t)$ and $\textbf{B}(t)$ to each other
\begin{equation}\label{300}
\fl \textbf{A}(t)=\hat{G}(t)\textbf{B}(t) \ \ \rmor \quad \sum_{n}c_{n}\{\alpha\hat{\cal E}_{1}+\beta\hat{\cal E}_{2}\}\ |\varphi_{n}\rangle=\hat{G}(t){\sum_{n}c_{n}\{\alpha\hat{\cal E}_{3}+\beta\hat{\cal E}_{4}\}\ |\varphi_{n}\rangle}.
\end{equation}
Now, for the global state of the system and the environment, which is given by
\begin{eqnarray}\label{310}
\nonumber |\Psi^{\rm tot}(t)\rangle=\textbf{A}(t)\ |a\rangle+\textbf{B}(t)\ |b\rangle=\hat{G}(t)\textbf{B}(t)\ |a\rangle+\textbf{B}(t)\ |b\rangle\\=\{\hat{G}(t)|a\rangle+|b\rangle\}\times(\sum_{n=0}^{\infty}c_{n}\{\alpha\hat{\cal E}_{3}+\beta\hat{\cal E}_{4}\}\ |\varphi_{n}\rangle),
\end{eqnarray}
we observe that if for some initial states of the system and the environment $\hat{G}(t)$ turns out to become in the form
\begin{eqnarray}\label{30.1}
\hat{G}(t)=G(t)\times\hat{I}_{\cal E},
\end{eqnarray}
with $G(t)$ as a scalar (rather than an operator) and $\hat{I}_{\cal E}$ representing the identity operator in the Hilbert space of the environment, then those initial states of the system and the environment will not entangle with each other, and hence they can represent the initial pointer states of the system and the environment. This result simply is because of the fact that if for some initial states of the system and the environment $\hat{G}(t)$ turns out to become a scalar in the form of equation (\ref{30.1}), $G(t)$ will be independent of the indices of the environment (i.e.\ independent of $n$); as in this case all components of $\textbf{B}(t)$ will be mapped into their corresponding components from $\textbf{A}(t)$ through the \emph{same} scalar function $G(t)$ (which will keep the two vectors $\textbf{A}(t)$ and $\textbf{B}(t)$ parallel to each other). Therefore, in this case $\hat{G}(t)$ will not enter the summation in the expression $\sum_{n}c_{n}\{\alpha\hat{\cal E}_{3}+\beta\hat{\cal E}_{4}\}\ |\varphi_{n}\rangle$ of equation (\ref{310}); and (as one can see from equation (\ref{310})) the states of the system and the environment respectively represented by $\{G(t)|a\rangle+|b\rangle\}$ and $\sum_{n}c_{n}\{\alpha\hat{\cal E}_{3}+\beta\hat{\cal E}_{4}\}\ |\varphi_{n}\rangle$ will not entangle to each other.

In another word, if for some initial states of the system and the environment $\textbf{A}(t)=\hat{G}(t)\textbf{B}(t)$ is equal to $G\textbf{B}(t)$, it means that for those initial states of the system and the environment $\textbf{B}(t)$ becomes an eigenstate of the operator $\hat{G}(t)$; and the two vectors $\textbf{A}(t)$ and $\textbf{B}(t)$ will stay parallel with each other throughout their evolution with time; and as we discussed, in this case the states of the system and the environment will not entangle with each other and (as one can see from equation (\ref{310})) pointer states can be realized for the system and the environment given by
\begin{eqnarray}\label{320}
\nonumber |\pm(t)\rangle=\calN_{\pm}\ \{G(t)|a\rangle+|b\rangle\} \qquad \rmand \\
|\Phi_{\pm}(t)\rangle=\calN_{\pm}^{-1}(\sum_{n=0}^{\infty}c_{n}\{\alpha\hat{\cal E}_{3}+\beta\hat{\cal E}_{4}\}\ |\varphi_{n}\rangle).
\end{eqnarray}
In the above equation we have represented the pointer states of the system by $|\pm(t)\rangle$ and those of the environment by $|\Phi_{\pm}(t)\rangle$. Also, $\calN_{\pm}$ is the normalization factor for the pointer states of the system (clearly $\calN_{\pm}=\rm\frac{1}{\sqrt{2}}$ if $|G(t)|=1$, as for the example of the JCM in the exact-resonance regime).

The condition represented by equation (\ref{30.1}) in fact is the \emph{necessary} condition for obtaining pointer states; since unless $\hat{G}(t)$ turns out to become a scalar, the two vectors $\textbf{A}(t)=\hat{G}(t)\times\textbf{B}(t)$ and $\textbf{B}(t)$ will not be parallel at all times and the operator $\hat{G}(t)$ will enter the summation over the environmental degrees of freedom (i.e.\ the summation over $n$) in equation (\ref{310}), in which case no longer the states of the system and the environment will be separable in a tensor product form; and pointer states cannot be realized for the states of the system and the environment. Also, as we discussed in \cite{paper3}, generally there is no guaranty for the condition (\ref{30.1}) to be satisfied; and satisfaction of this condition often may require having a sufficiently large environment which contains a large number of
degrees of freedom. However, \emph{if} in some regime and for a given Hamiltonian defining a system-environment model we can find initial states for the system and the environment which satisfy this condition, we do know that pointer states can be realized for the system and the environment and those initial states would correspond to the initial pointer states of the system and the environment.

In essence, in order to find the pointer states of the system and the environment for a given total Hamiltonian defining our system-environment model, and for a given initial state of the environment, our main goal would be finding those possible initial states of the system for which $\hat{G}(t)$ (which is defined through equation (\ref{300})) is of the form of equation (\ref{30.1}). In section 3 considering the quantized atom-field model represented by the Hamiltonian of equation (\ref{1}) and for the regime that $\hat{H}_{\cal E}\ll\hat{H}_{\cal S}\ll\hat{H'}$ (but $\hat{H}_{\cal E}\neq0$ and $\hat{H}_{\cal S}\neq0$), we exploit this method to obtain the time-dependent pointer states of the system and the environment; by assuming an initial state of the environment in the form of a Gaussian package in position space. As we will see, once we have the time-evolution operator for our system-environment model in the form of equation (\ref{270}) and the $\hat{\cal E}_{i}$ operators, this task can be done quite easily for our model.\\

\section{Calculation of the time-evolution operator}

In order to calculate the time-evolution operator in the interaction picture for the Hamiltonian of equation (\ref{1}), first we need to have the Hamiltonian in the interaction picture, which is defined through the following equation
\begin{equation}\label{11}
\hat{H}_{\rmint}(t)=e^{i\hat{H}_{0}t}\hat{H}^{'}e^{-i\hat{H}_{0}t}.
\end{equation}
Here $\hat{H}_{0}=\frac{1}{2}\omega_{0}\hat{\sigma}_{z}+\omega\hat{a}^{\dag}\hat{a}$ is the sum of the self Hamiltonians of the system and the environment; and $\hat{H}^{'}=\rmg \chi \hat{\sigma}_{x} \hat{x}$ is the Hamiltonian for the interaction between the system and the environment.
So, now we must calculate
\begin{equation}\label{12}
\hat{H}_{\rmint}(t)=\rmg\ (e^{i\omega_{0}\hat{\sigma}_{z}t/2}\ \hat{\sigma}_{x}\ e^{-i\omega_{0}\hat{\sigma}_{z}t/2})\otimes(e^{i\omega\hat{a}^{\dag}\hat{a}t}\ \chi\hat{x}\ e^{-i\omega\hat{a}^{\dag}\hat{a}t}),
\end{equation}
where $\chi\hat{x}=\hat{a}+\hat{a}^{\dag}$. However, $\hat{\sigma}_{x}=\hat{\sigma}_{+}+\hat{\sigma}_{-}$; and $e^{i\omega_{0}\hat{\sigma}_{z}t/2}\ \hat{\sigma}_{\pm}\ e^{-i\omega_{0}\hat{\sigma}_{z}t/2}=\hat{\sigma}_{\pm}\ e^{\pm i\omega_{0}t}$. Also $\hat{a}(t)=\hat{a}e^{-i\omega t}$. So
\begin{eqnarray}\label{13}
\nonumber \hat{H}_{\rmint}(t)=\rmg(\hat{\sigma}_{+}\ e^{i\omega_{0}t}+\hat{\sigma}_{-}\ e^{-i\omega_{0}t})\otimes(\hat{a}e^{-i\omega t}+\hat{a}^{\dag}e^{i\omega t})\\=\rmg\{\hat{\sigma}_{+}(\hat{a}\ e^{i\Delta t}+\hat{a}^{\dag}\ e^{i(\omega+\omega_{0}) t})+c.c.\}, \qquad \rmwith \qquad \Delta=\omega_{0}-\omega.
\end{eqnarray}

Now in parallel with \cite{paper1}, for the evolution operator of the global composite system we consider the general form given by equation (\ref{270}). For such time-evolution operator in the interaction picture, which satisfies the Schr\"{o}dinger equation
\begin{equation}\label{14}
i\frac{\partial}{\partial t}\hat{U}(t)=\hat{H}_{\rm int}\hat{U}(t),
\end{equation}
we have
\begin{eqnarray}\label{15}
\nonumber i \left(
         \begin{array}{cc}
           \dot{\hat{\cal E}_{1}} & \dot{\hat{\cal E}_{2}} \\
           \dot{\hat{\cal E}_{3}} & \dot{\hat{\cal E}_{4}} \\
         \end{array}
       \right)=\hat{H}_{\rmint}(t)\left(
                 \begin{array}{cc}
                   \hat{\cal E}_{1} & \hat{\cal E}_{2} \\
                   \hat{\cal E}_{3} & \hat{\cal E}_{4} \\
                 \end{array}
               \right)\\=\rmg\left(
                 \begin{array}{cc}
                   0& \hat{a}\ e^{i\Delta t}+\hat{a}^{\dag}\ e^{i(\omega+\omega_{0}) t} \\
                   \hat{a}^{\dag}\ e^{-i\Delta t}+\hat{a}\ e^{-i(\omega+\omega_{0}) t} & 0 \\
                 \end{array}
               \right)\left(
                 \begin{array}{cc}
                   \hat{\cal E}_{1} & \hat{\cal E}_{2} \\
                   \hat{\cal E}_{3} & \hat{\cal E}_{4} \\
                 \end{array}
               \right)\\ \nonumber =\rmg\left(
                 \begin{array}{cc}
                   (\hat{a}\ e^{i\Delta t}+\hat{a}^{\dag}\ e^{i(\omega+\omega_{0}) t})\ \hat{\cal E}_{3} & (\hat{a}\ e^{i\Delta t}+\hat{a}^{\dag}\ e^{i(\omega+\omega_{0}) t})\ \hat{\cal E}_{4} \\
                   (\hat{a}^{\dag}\ e^{-i\Delta t}+\hat{a}\ e^{-i(\omega+\omega_{0}) t})\ \hat{\cal E}_{1} & (\hat{a}^{\dag}\ e^{-i\Delta t}+\hat{a}\ e^{-i(\omega+\omega_{0}) t})\ \hat{\cal E}_{2} \\
                 \end{array}
               \right).
\end{eqnarray}

Now, we assume $\omega\ll\omega_{0}$; so that $\Delta\approx\omega_{0}$ and $\omega+\omega_{0}\approx\omega_{0}$. In other words, in the Hamiltonian of our total composite system, given by equation (\ref{1}), we assume that the self-Hamiltonian of the system dominates the self-Hamiltonian of the environment. Therefore, equation (\ref{15}) for the evolution of the time-evolution operator can be simplified to the following set of four equations
\begin{eqnarray}\label{16}
\nonumber i\dot{\hat{\cal E}_{1}}=\rmg \chi\hat{x}\ e^{i\omega_{0}t}\ \hat{\cal E}_{3},\\
\nonumber i\dot{\hat{\cal E}_{2}}=\rmg \chi\hat{x}\ e^{i\omega_{0}t}\ \hat{\cal E}_{4},\\
i\dot{\hat{\cal E}_{3}}=\rmg\chi\hat{x}\ e^{-i\omega_{0}t}\ \hat{\cal E}_{1},\\
\nonumber i\dot{\hat{\cal E}_{4}}=\rmg \chi\hat{x}\ e^{-i\omega_{0}t}\ \hat{\cal E}_{2}.
\end{eqnarray}

In order to solve the above set of coupled differential equations, we proceed as follows. First, we take derivative with respect to time of the first equation. By replacing $\dot{\hat{\cal E}_{3}}$ from the third equation in the resulting equation we find
\begin{equation}\label{17}
\ddot{\hat{\cal E}_{1}}=-(\rmg \chi\hat{x})^{2}\ \hat{\cal E}_{1}+(\rmg \chi\hat{x}\omega_{0}\ e^{i\omega_{0}t})\ \hat{\cal E}_{3}.
\end{equation}
Similarly, by doing the same procedure on the third equation for $\dot{\hat{\cal E}_{3}}$ we find
\begin{equation}\label{18}
\ddot{\hat{\cal E}_{3}}=-(\rmg \chi\hat{x})^{2}\ \hat{\cal E}_{3}-(\rmg \chi\hat{x}\omega_{0}\ e^{-i\omega_{0}t})\ \hat{\cal E}_{1}.
\end{equation}
One can easily verify that \emph{if} $\omega_{0}^{2}\ll(\rmg\chi)^{2}$ (i.e.\ \emph{if} $\hat{H}_{\cal S}\ll\hat{H'}$), so that $(\rmg \chi\hat{x})^{2}+\omega_{0}^{2}/4\approx(\rmg \chi\hat{x})^{2}$, the following solutions will satisfy the differential equations given by equations (\ref{17}) and (\ref{18}) for $\hat{\cal E}_{1}$ and $\hat{\cal E}_{3}$:
\begin{equation}\label{19}
\hat{\cal E}_{1}=\cos(\rmg \chi\hat{x}t)\ e^{i\omega_{0}t/2} \qquad \rmand \qquad \hat{\cal E}_{3}=-i\sin(\rmg \chi\hat{x}t)\ e^{-i\omega_{0}t/2}.
\end{equation}
In quite the same manner we can calculate $\hat{\cal E}_{2}$ and $\hat{\cal E}_{4}$ as follows
\begin{equation}\label{20}
\hat{\cal E}_{2}=-i\sin(\rmg \chi\hat{x}t)\ e^{i\omega_{0}t/2} \qquad \rmand \qquad \hat{\cal E}_{4}=\cos(\rmg \chi\hat{x}t)\ e^{-i\omega_{0}t/2}.
\end{equation}

The above operators together with equation (\ref{270}) make the time-evolution operator of the quantized atom-field model and for the regime that $\hat{H}_{\cal E}\ll\hat{H}_{\cal S}\ll\hat{H'}$, but $\hat{H}_{\cal S}\neq0$ and $\hat{H}_{\cal E}\neq0$. One can easily verify that the above set of operators satisfies the unitarity of the time-evolution operator given by $\hat{U}^{\dag}\hat{U}=\hat{U}\hat{U}^{\dag}=\hat{I}$ (with $\hat{I}$ representing the identity operator). Moreover, $\hat{\cal E}_{1}(0)=\hat{\cal E}_{4}(0)=1$ and $\hat{\cal E}_{2}(0)=\hat{\cal E}_{3}(0)=0$. So, these operators do satisfy the initial condition for the time-evolution operator given by $\hat{U}_{\rmtot}(t_0)=\hat{I}$.

\section{Calculation of the time-dependent pointer states of the system and the environment}
Using the time-evolution operator which we already obtained for our model and for the regime that $\hat{H}_{\cal E}\ll\hat{H}_{\cal S}\ll\hat{H'}$ (but $\hat{H}_{\cal S}\neq0$ and $\hat{H}_{\cal E}\neq0$), now we can obtain the corresponding pointer states of the system and the environment in this regime. For this purpose we assume that the system initially is prepared in the state $|\psi^{\cal S}(t_{0})\rangle=\alpha|a\rangle+\beta|b\rangle$. Moreover, let us assume that the initial state of the environment can be represented by a Gaussian package in the position space
\begin{equation}\label{21}
|\Phi^{\cal E}(t_{0})\rangle=\calN_{0}\int_{-\infty}^{\infty}dx\ e^{-\alpha_{\circ} x^{2}}|x\rangle,
\end{equation}
where $\calN_{0}=(2\alpha_{\circ}/\pi)^{1/4}$ is the normalization factor for this state. Now, the condition for determining the pointer states of the system and the environment, given by equations (\ref{300}) and (\ref{30.1}), reads
\begin{eqnarray}\label{22}
\nonumber (\alpha\hat{\cal E}_{1}+\beta\hat{\cal E}_{2})\ |\Phi^{\cal E}(t_{0})\rangle=\hat{G}(t)\times(\alpha\hat{\cal E}_{3}+\beta\hat{\cal E}_{4})
\ |\Phi^{\cal E}(t_{0})\rangle; \\ with\ \hat{G}(t)\ being\ proportional\ to\ the\ unit\ matrix.
\end{eqnarray}
(In other words, for an initial state of the system corresponding to one of its pointer states at $t=t_{0}$, the operator $\hat{G}(t)$ must be independent of the indices of the environment. i.e.\ $x$). Inserting the $\hat{\cal E}_{i}$'s from equations (\ref{19}) and (\ref{20}) into the above condition it reads
\begin{eqnarray}\label{24}
\nonumber \int_{-\infty}^{\infty}dx\ [\alpha\cos(\rmg \chi xt)-i\beta\sin(\rmg \chi xt)]\ e^{-\alpha_{\circ}x^{2}+i\omega_{0}t/2}|x\rangle\\ =\hat{G}(t)\times\int_{-\infty}^{\infty}dx\ [-i\alpha\sin(\rmg \chi xt)+\beta\cos(\rmg \chi xt)]\ e^{-\alpha_{\circ}x^{2}-i\omega_{0}t/2}|x\rangle\\ \nonumber and\ \hat{G}(t)\ be\ proportional\ to\ the\ unit\ matrix.
\end{eqnarray}
For pointer states $\hat{G}(t)$ must satisfy the condition (\ref{30.1}) for obtaining the pointer states of the system and the environment, i.e.\ $\hat{G}(t)=G(t)\times\hat{I}_{\cal E}$. Therefore, since the set $\{|x\rangle\}$ is a complete set of basis states for the environment, for initial pointer states we can simply equalize those terms from the two sides of equation (\ref{24}) which correspond to the same $|x\rangle$ state and obtain
\begin{equation}\label{25}
G(t)=\frac{\alpha\cos(\rmg \chi xt)-i\beta\sin(\rmg \chi xt)}{-i\alpha\sin(\rmg \chi xt)+\beta\cos(\rmg \chi xt)}\ e^{i\omega_{0}t}.
\end{equation}
The above result for $G(t)$, which generally depends on $x$, would contradict our initial assumption of $\hat{G}(t)$ being proportional to the unit matrix \emph{unless} if we can find certain initial states for the system for which $G(t)$ turns out to become independent of $x$; since, as we discussed, for pointer states, all components of the vector $\textbf{A}$ (${A_{x}}^{'}s$) must be related to their corresponding components from $\textbf{B}$ (${B_{x}}^{'}s$) through the \emph{same} scalar factor $G$ (see equations (\ref{300}) and (\ref{30.1})).\footnote{We would like to see if the condition can be satisfied for \emph{any} initial state of the system and the environment with $G(t)$ becoming independent of the variable $x$ of the states of the environment. So, if finally we can find any specific set of initial states for the system and the environment which satisfies this condition with $G(t)$ independent of the indices of the environment, then we have reached our goal.} So now we should seek for those particular initial states of the system which can make $G(t)$ independent of the variable $x$ of the states of the environment.

From equation (\ref{25}) we easily see that for $\alpha=\pm \beta$, $G(t)$ turns out to become
\begin{equation}\label{26}
G(t)=\pm e^{i\omega_{0} t}
\end{equation}
which clearly is independent of the variable $x$ of the states of the environment.

The above result simply means that for the initial states of the system obtained from
\begin{equation}\label{27}
\fl \alpha_{+}=\beta_{+}=\frac{1}{\sqrt{2}}\  \quad \rmand \quad \alpha_{-}=-\beta_{-}=\frac{1}{\sqrt{2}}\ , \quad \rmor \quad |\pm(t_{0})\rangle=\frac{1}{\sqrt{2}}(|a\rangle\pm|b\rangle),
\end{equation}
(which correspond to the initial conditions for the state of the system given by $\alpha=\pm \beta$) the states of the system and the environment will not entangle with each other. Moreover, using equation (\ref{320}), which gives us the general time evolution of the pointer states of the system, and $G(t)$ of equation (\ref{26}) (which is independent of the variable $x$ of the states of the environment) we can find the time evolution of the pointer states of the system as follows
\begin{eqnarray}\label{29}
|\pm(t)\rangle=\calN\ \{G(t)|a\rangle+|b\rangle\}=\frac{1}{\sqrt{2}}(e^{i\omega_{0} t}\ |a\rangle\pm|b\rangle).
\end{eqnarray}
As we observe, in the regime that we are considering ($\hat{H}_{\cal E}\ll\hat{H}_{\cal S}\ll\hat{H'}$, with $\hat{H}_{\cal E}\neq0$ and $\hat{H}_{\cal S}\neq0$), $G(t)$ and the time evolution of the pointer states of the system are characterized by $\omega_{0}$ of the self-Hamiltonian of the system; unlike the exact-resonance with the rotating wave approximation regime where the evolution of the pointer states of the system is characterized by the atom-field coupling constant $\rmg$ and the average number of photons $\bar{n}$, through the factor $\rmg/\sqrt{\bar{n}}$ \cite{Gea-Banacloche,Gea-Banacloche2}.

Next, we obtain the corresponding pointer states of the environment. Using equations (\ref{320}), (\ref{21}) and (\ref{27}) we have
\begin{eqnarray}\label{30}
\nonumber |\Phi_{\pm}(t)\rangle=\calN^{-1}(\alpha_{\pm}\hat{\cal E}_{3}+\beta_{\pm}\hat{\cal E}_{4})\ |\Phi^{\cal E}(t_{0})\rangle \\ =
\calN_{0}(\hat{\cal E}_{3}\pm\hat{\cal E}_{4})\ \int_{-\infty}^{\infty}dx\ e^{-\alpha_{\circ}x^{2}}|x\rangle;
\end{eqnarray}
since $\calN^{-1}\alpha_{\pm}=1$ and $\calN^{-1}\beta_{\pm}=\pm 1$. Therefore,
\begin{eqnarray}\label{31}
|\Phi_{\pm}(t)\rangle=(\frac{2\alpha_{\circ}}{\pi})^{\frac{1}{4}}\ \int_{-\infty}^{\infty}dx\ e^{-\alpha_{\circ}x^{2}\mp i(\rmg \chi x\pm\omega_{0}/2)t}\ |x\rangle.
\end{eqnarray}
Also, the overlap between the pointer states of the environment can be calculated as
\begin{equation}\label{31.1}
\langle\Phi_{-}(t)|\Phi_{+}(t)\rangle=e^{-(\rmg\chi t)^{2}/2\alpha_{\circ}}.
\end{equation}

We should also mention that the pointer states of the system at $t=t_{0}$ (see equation (\ref{27})) are orthonormal and hence, they form a complete basis set for the state of the system. Therefore, the evolution of any initial pure state of the two-level system $|\psi_{\cal S}(t_{0})\rangle=\alpha'\ |+(t_{0})\rangle+\beta'\ |-(t_{0})\rangle$ with an initial field $|\Phi_{\cal E}(t_{0})\rangle$, in the form of equation (\ref{21}), can be expressed as a linear combination of the evolution of $|+(t_{0})\rangle|\Phi_{\cal E}(t_{0})\rangle$ and
$|-(t_{0})\rangle|\Phi_{\cal E}(t_{0})\rangle$
\begin{eqnarray}\label{32}
\fl (\alpha'\ |+(t_{0})\rangle+\beta'\ |-(t_{0})\rangle)\ |\Phi_{\cal E}(t_{0})\rangle\rightarrow
    \alpha'\ |+(t)\rangle\ |\Phi_{+}(t)\rangle+\beta'\ |-(t)\rangle\ |\Phi_{-}(t)\rangle,
\end{eqnarray}
where in the above equation the evolution of the pointer states of the system $|\pm(t)\rangle$ is given by equation (\ref{29}) and the evolution of the pointer states of the environment $|\Phi_{\pm}(t)\rangle$ is given by equation (\ref{31}).

\section{Consequences regarding the decoherence of the central system}

In this section first we use the time-evolution operator, already obtained in section 2, to obtain the general time evolution of the total composite system for our model and for an initial state of the environment in the form of a Gaussian package in position space, such as that of equation (\ref{21}). After that, we will calculate the offdiagonal element of the reduced density matrix of the system (i.e.\ $\rho_{12}^{(\cal S)}(t)$) by tracing over the environmental degrees of freedom. Then, we will also obtain the coherences of the reduced density matrix of the system in another way by using the pointer states of the system and the environment obtained in section 3.
As we will see, the two results will be in perfect agreement with each other.
Finally, we will discuss some interesting features which can be observed in our study of the decoherence of the central system.

Using equations (\ref{290}), (\ref{19}) and (\ref{20}) to obtain $|\Psi_{\rmtot}(t)\rangle$, we can write
\begin{eqnarray}\label{33}
\fl \nonumber |\Psi_{\rmtot}(t)\rangle=\textbf{A}(t)\ |a\rangle+\textbf{B}(t)\ |b\rangle=(\alpha\cos(\rmg \chi\hat{x}t)\ e^{i\omega_{0}t/2}-i\beta\sin(\rmg \chi\hat{x}t)\ e^{i\omega_{0}t/2})\ |\Phi^{\cal E}(t_{0})\rangle|a\rangle\\+(-i\alpha\sin(\rmg \chi\hat{x}t)\ e^{-i\omega_{0}t/2}+\beta\cos(\rmg \chi\hat{x}t)\ e^{-i\omega_{0}t/2})\ |\Phi^{\cal E}(t_{0})\rangle|b\rangle.
\end{eqnarray}
In the above equation $|\Phi^{\cal E}(t_{0})\rangle$ is the initial state of the environment, represented by the Gaussian package of equation (\ref{21}).

For the state of the total composite system in our model, which is given by equation (\ref{33}), we can do the trace operation over the basis states of the environment (i.e.\ the $\{|x\rangle\}$ which make a complete basis for the state of the environment) to obtain the reduced  density matrix of the system $\cal S$
\begin{eqnarray}\label{34}
\nonumber \hat{\rho}_{\cal S}(t)=\int_{-\infty}^{\infty}dx\ \langle x|\hat{\rho}^{\rmtot}(t)|x\rangle=\int_{-\infty}^{\infty}dx\ \langle x|\Psi_{\rmtot}(t)\rangle\langle\Psi_{\rmtot}(t)|x\rangle\\ \fl =\int_{-\infty}^{\infty}dx\ (\ |\psi_{a}(x,t)|^{2}\ |a\rangle\langle a|+|\psi_{b}(x,t)|^{2}\ |b\rangle\langle b|+\psi_{a}(x,t)\psi_{b}^{\ast}(x,t)\ |a\rangle\langle b|+c.c.\ ).
\end{eqnarray}
where
\begin{eqnarray}\label{35}
\nonumber \psi_{a}(x,t)=(\frac{2\alpha_{\circ}}{\pi})^{\frac{1}{4}}\ [\alpha\cos(\rmg \chi xt)\ e^{i\omega_{0}t/2}-i\beta\sin(\rmg \chi xt)\ e^{i\omega_{0}t/2}]\ e^{-\alpha_{\circ}x^{2}} \quad \rmand \\ \psi_{b}(x,t)=(\frac{2\alpha_{\circ}}{\pi})^{\frac{1}{4}}\ [-i\alpha\sin(\rmg \chi xt)\ e^{-i\omega_{0}t/2}+\beta\cos(\rmg \chi xt)\ e^{-i\omega_{0}t/2}]\ e^{-\alpha_{\circ}x^{2}}.
\end{eqnarray}
Using equations (\ref{34}) and (\ref{35}), after doing the integrations we easily find
\begin{eqnarray}\label{36}
\fl \nonumber \rho^{\cal S}_{aa}(t)=1-\rho^{\cal S}_{bb}(t)=\int_{-\infty}^{\infty}dx\ |\psi_{a}(x,t)|^{2}=\frac{1}{2}[1+(|\alpha|^2-|\beta|^2)\ e^{-(\rmg\chi t)^{2}/2\alpha_{\circ}}]\quad \rmand \\ \fl
\rho^{\cal S}_{ab}(t)=\int_{-\infty}^{\infty}dx\ \psi_{a}(x,t)\psi_{b}^{\ast}(x,t)= \frac{1}{2}[(\alpha\beta^{\ast}+\beta\alpha^{\ast})+ (\alpha\beta^{\ast}-\beta\alpha^{\ast})e^{-(\rmg\chi t)^{2}/2\alpha_{\circ}}]e^{i\omega_{0}t}.
\end{eqnarray}
(In the above equation we used the notation $\rho_{ab}=\langle a|\hat{\rho}_{\cal S}(t)|b\rangle$ and etc.) As we see from the above equations, for the initial pointer states of the system, for which $|\alpha|=|\beta|$ (see equation (\ref{27})), and also for very large times $t\rightarrow\infty$, the diagonal elements of the reduced density matrix of the system will be equal to the constant number of $\frac{1}{2}$. Also, for the initial pointer states of the system
we have $\rho^{\cal S}_{ab}(t)=\frac{1}{2}(\alpha\beta^{\ast}+\beta\alpha^{\ast})e^{i\omega_{0}t}$. This means that for the initial pointer states of the system $|\rho^{\cal S}_{ab}(t)|$ will always be equal to the constant value of $\frac{1}{2}$; while for most of the other states (for whom $\alpha\beta^{\ast}\neq\beta\alpha^{\ast}$) only at sufficiently large times $|\rho^{\cal S}_{ab}(t)|$ will converge to the constant value of $\frac{1}{2}(\alpha\beta^{\ast}+\beta\alpha^{\ast})$, with a decoherence time given by
\begin{equation}\label{37}
\tau_{\rm dec}=\frac{\hbar}{\rmg}\frac{\sqrt{2\alpha_{\circ}}}{\chi}=\frac{\hbar}{\rmg}\sqrt{\frac{\alpha_{\circ}}{m\omega}}.
\end{equation}

The reduced density matrix of a two-level system $\hat{\rho}_{\cal S}(t)$ generally can be expressed in terms of the Bloch vector $\textbf{R}(t)\equiv(R_{x}, R_{y}, R_{z})$ \cite{Eberly} as follows
\begin{equation}\label{38}
\hat{\rho}_{\cal S}(t)=\frac{1}{2}(\hat{I}+\textbf{R}(t).\hat{\mathbf{\sigma}})=\frac{1}{2}(\hat{I}+R_{x}\sigma_{x}+R_{y}\sigma_{y}+R_{z}\sigma_{z});
\end{equation}
from which one can easily verify that the Bloch vector components must be defined by
\begin{equation}\label{39}
R_{x}=\rho_{ab}+\rho_{ba} \qquad R_{y}=i(\rho_{ab}-\rho_{ba})\quad \rmand \quad R_{z}=\rho_{aa}-\rho_{bb}.
\end{equation}
So now, using our expressions for the elements of the reduced density matrix of the system, given by equation (\ref{36}), we can also calculate the components of the Bloch vector, which are a measure for the polarization of the state of the two-level system \cite{paper1,Schlosshauer}. One would easily find
\begin{eqnarray}\label{40}
\fl \nonumber R_{x}(t)=\rho_{ab}+\rho_{ab}^{\ast}=(\alpha\beta^{\ast}+\beta\alpha^{\ast})\cos(\omega_{0}t)+i(\alpha\beta^{\ast}-\beta\alpha^{\ast})\sin(\omega_{0}t)\  e^{-(\rmg\chi t)^{2}/2\alpha_{\circ}}, \\ \fl
R_{y}(t)=i(\rho_{ab}-\rho_{ab}^{\ast})=-(\alpha\beta^{\ast}+\beta\alpha^{\ast})\sin(\omega_{0}t)+i(\alpha\beta^{\ast}-\beta\alpha^{\ast})\cos(\omega_{0}t)\  e^{-(\rmg\chi t)^{2}/2\alpha_{\circ}}, \\ \fl
\nonumber R_{z}(t)=\rho_{aa}-\rho_{bb}=(|\alpha|^2-|\beta|^2)\ e^{-(\rmg\chi t)^{2}/2\alpha_{\circ}}.
\end{eqnarray}
For $t\rightarrow\infty$ and $\chi\neq0$ we have
\begin{eqnarray}\label{41}
\nonumber R_{x}(t)\rightarrow(\alpha\beta^{\ast}+\beta\alpha^{\ast})\cos(\omega_{0}t), \\
R_{y}(t)\rightarrow -(\alpha\beta^{\ast}+\beta\alpha^{\ast})\sin(\omega_{0}t)  \quad \rmand \\
\nonumber R_{z}(t)\rightarrow 0.
\end{eqnarray}
The above result simply means that at $t\rightarrow\infty$ and if $\chi=\sqrt{2m\omega}\neq0$ the pointer states of the system will evolve between the eigenstates of the $\hat{\sigma}_{x}$ and $\hat{\sigma}_{y}$ Pauli matrices; and therefore, a preferred basis of measurement is \emph{not} determined in the regime that we are considering; although the eigenstates of $\hat{\sigma}_{z}$ are excluded from being realized at $t\rightarrow\infty$.

One can easily obtain the coherences of the reduced density matrix of the system in another way by using the pointer states of the system and the environment which we obtained in section 3. As we saw, for a two-state system $\cal S$ in contact with an environment $\cal E$ after determination of the pointer states of the system and the environment, the state of the total composite system generally can be represented by equation (\ref{32}). i.e.\ $|\Psi_{\rmtot}(t)\rangle=\alpha'\ |+(t)\rangle\ |\Phi_{+}(t)\rangle+\beta'\ |-(t)\rangle\ |\Phi_{-}(t)\rangle$. For $|\Psi_{\rmtot}(t)\rangle$ given by equation (\ref{32}) the reduced density matrix of the system $\hat{\rho}_{\cal S}(t)$ can be calculated by tracing over the environmental degrees of freedom to obtain
\begin{eqnarray}\label{42}
\nonumber \hat{\rho}_{\cal S}(t)=|\alpha'|^{2}\times|+(t)\rangle\langle +(t)|+|\beta'|^{2}\times|-(t)\rangle\langle -(t)|+\alpha'\beta'^{\ast}\\ \fl \times|+(t)\rangle\langle -(t)|\times\langle\Phi_{-}(t)|\Phi_{+}(t)\rangle+\beta'\alpha'^{\ast}\times|-(t)\rangle\langle +(t)|\times\langle\Phi_{+}(t)|\Phi_{-}(t)\rangle.
\end{eqnarray}
So, in an arbitrary basis $|a\rangle$ and $|b\rangle$ of the state of the two-level system generally we have
\begin{eqnarray}\label{43}
\fl \nonumber \rho^{\cal S}_{aa}(t)=1-\rho^{\cal S}_{bb}(t)=|\alpha'|^{2}\times\langle a|+(t)\rangle\langle +(t)|a\rangle+|\beta'|^{2}\times\langle a|-(t)\rangle\langle -(t)|a\rangle+\alpha'\beta'^{\ast}\\ \fl \times\langle a|+(t)\rangle\langle -(t)|a\rangle\times\langle\Phi_{-}(t)|\Phi_{+}(t)\rangle+\beta'\alpha'^{\ast}\times\langle a|-(t)\rangle\langle +(t)|a\rangle\times\langle\Phi_{+}(t)|\Phi_{-}(t)\rangle \\ \fl \nonumber \rmand \qquad
\rho^{\cal S}_{ab}(t)=|\alpha'|^{2}\times\langle a|+(t)\rangle\langle +(t)|b\rangle+|\beta'|^{2}\times\langle a|-(t)\rangle\langle -(t)|b\rangle+\alpha'\beta'^{\ast}\\ \fl \nonumber \times\langle a|+(t)\rangle\langle -(t)|b\rangle\times\langle\Phi_{-}(t)|\Phi_{+}(t)\rangle+\beta'\alpha'^{\ast}\times\langle a|-(t)\rangle\langle +(t)|b\rangle\times\langle\Phi_{+}(t)|\Phi_{-}(t)\rangle.
\end{eqnarray}

The expansion coefficients $\alpha'$ and $\beta'$ for the state of the two-level system in the basis of the $|\pm(t_{0})\rangle$ states are related to the corresponding coefficients in the $|a\rangle$ and $|b\rangle$ basis\footnote{Now by $|a\rangle$ and $|b\rangle$ we mean the upper and lower levels of the two-level system; i.e.\ $|a\rangle$ and $|b\rangle$ no longer are some arbitrary basis states for the state of the two-level system.} through $\alpha'=\frac{1}{\sqrt{2}}(\alpha+\beta)$ and $\beta'=\frac{1}{\sqrt{2}}(\alpha-\beta)$. So now, for our quantized atom-field model and in the regime that we are considering one can use equations (\ref{29}) and (\ref{31.1}) to calculate the expressions in equation (\ref{43}) for the elements of the reduced density matrix of the system; obtaining exactly the same results as those of equation (\ref{36}).

One could similarly study the decoherence of the state of the system in the basis of the $|\pm(t_{0})\rangle$ states. As one can see from equation (\ref{29}), for $t\ll\omega_{0}^{-1}$ the pointer states of the system approximately can be represented by $|\pm(t_{0})\rangle$. Therefore, in the basis of the $|\pm(t_{0})\rangle$ states the short-time evolution of the off-diagonal element of the reduced density matrix of the system should be given by
\begin{equation}\label{45}
\rho^{\cal S}_{12}(t)\approx\alpha'\beta'^{\ast}\langle\Phi_{-}(t)|\Phi_{+}(t)\rangle=\alpha'\beta'^{\ast}\ e^{-(\rmg\chi t)^{2}/2\alpha_{\circ}}
\end{equation}

Hence, in the basis of the $|\pm(t_{0})\rangle$ states the short-time decoherence of the state of the system is characterized by the decaying factor
$e^{-(\rmg\chi t)^{2}/2\alpha_{\circ}}$, when the system initially is \emph{not} prepared in one of its pointer states $(\alpha'\beta'^{\ast}\neq0)$; while in this basis the pointer states of the system (for whom $\alpha'\beta'=0$) almost do not decohere within short times; and $\rho^{\cal S}_{12}(t)\approx0$ at all short times (i.e.\ for $t\ll\omega_{0}^{-1}$ for which $|\pm(t)\rangle\approx|\pm(t_{0})\rangle$).

Finally, let us study whether the short-time decay of $\rho^{\cal S}_{12}(t)$, given by equation (\ref{45}), might be reversible or not. As we will show here, the coherences of the reduced density matrix of the system, may revive at a later time. In such cases, of course we cannot have irreversible decoherence.

Using equation (\ref{43}) for the offdiagonal element of the reduced density matrix of the system and equations (29) and (32), we can calculate the all-time evolution of $\rho^{\cal S}_{12}(t)$ for the regime that we are considering and in the basis of the initial pointer states of the system $|\pm(t_{0})\rangle$ as follows
\begin{eqnarray}\label{6}
\nonumber \rho^{\cal S}_{12}(t)=(|\beta'|^{2}-|\alpha'|^{2})\times[\frac{i}{2}\ \sin(\omega_{0}t)]+\alpha'\beta'^{\ast}\times\cos^{2}(\omega_{0}t/2)\times e^{-(\rmg\chi t)^{2}/2\alpha_{\circ}}\\+\beta'\alpha'^{\ast}\times\sin^{2}(\omega_{0}t/2)\times e^{-(\rmg\chi t)^{2}/2\alpha_{\circ}};
\end{eqnarray}
which its short time evolution ($t\ll\omega_{0}^{-1}$) is the same as equation (\ref{45}).

Now, clearly for $t\rightarrow\infty$ we have
\begin{equation}\label{7}
\rho^{\cal S}_{12}(t)=(|\beta'|^{2}-|\alpha'|^{2})\times[\frac{i}{2}\ \sin(\omega_{0}t)].
\end{equation}
Therefore, except for $|\alpha'|=|\beta'|$, in the basis of the initial pointer states of the system and for $t\rightarrow\infty$ the offdiagonal element of the reduced density matrix of the system, $\hat{\rho}^{\cal S}_{12}$, will be oscillating with the frequency of $\omega_{0}$. As a result, we should note that the short-time decay, represented by equation (\ref{45}), can be reversible; as $\rho^{\cal S}_{12}(t)$ may revive at a later time.

\section{Summary and conclusions}
Considering the quantized atom-field model of quantum optics, we obtained the time-evolution operator for the regime that $\hat{H}_{\cal E}\ll\hat{H}_{\cal S}\ll\hat{H'}$ (but $\hat{H}_{\cal S}\neq0$ and $\hat{H}_{\cal E}\neq0$). Using this time-evolution operator then we calculated the corresponding pointer states of the system and the environment, which are characterized by their ability not to entangle with each other, by assuming an initial state of the environment in the form of a Gaussian package in position space. Most importantly, we observed that for our model represented by the Hamiltonian of equation (\ref{1}) the pointer states of the system turn out to become \emph{time-dependent}, as opposed to the pointer states of some simpler models, which often are cited in the context of quantum information and quantum computation
[8-15]. However, in most of the practical situations different noncommutable perturbations may exist in the total Hamiltonian of a realistic system-environment model, which would result in having time-dependent pointer states for the system \cite{paper1}. Indeed, the authors believe that the fact that the pointer states of a system generally are time-dependent and may evolve with time has not been seriously acknowledged in the context of quantum computation and quantum information. In specific, in the context of quantum error correction \cite{Laflamme,Nielsen} it is often assumed that the premeasurement by the environment does not change the initial pointer states of the system. In other words, quantum ``nondemolition" premeasurement by the environment often is assumed \cite{Laflamme,Nielsen}; as is also assumed in the Von Neumann scheme of measurement \cite{Neuman,Schlosshauer}. Also, in the context of Decoherence-Free-Subspaces (DFS) theory the models which often are studied either contain self-Hamiltonians for the system which commute with the interaction between the system and the environment, or it is assumed that we are in the \emph{quantum measurement limit} \footnote{In the \emph{quantum measurement limit} the interaction between the system and the environment is so strong as to dominate the evolution of the system $\hat{H}\approx \hat{H}_{\rm int}$. Also in the \emph{quantum limit of decoherence} the Hamiltonian for the system almost dominates the interaction between the system and the environment as well as the self-Hamiltonian of the environment $\hat{H}\approx \hat{H}_{\cal S}$.} or in the \emph{quantum limit of decoherence} \cite{Ekert,90,91,93}. However, all of these assumptions are in fact a simplification of the problem; since, as we discussed in \cite{paper1}, they completely exclude the possibility of having pointer states for the system which may depend on time.

Using the time-evolution operator obtained in section 2, we also obtained a closed form for the elements of the reduced density matrix of the system, and studied the decoherence of the central system in our model. We showed that for the case that the system initially is not prepared in one of its pointer states and in the basis of the initial pointer states of the system (i.e.\ the $|\pm(t_{0})\rangle $ states), the short time ($t\ll\omega_{0}^{-1}$) evolution of the offdiagonal elements of the reduced density matrix of the system will demonstrate decoherence, with a decoherence factor given by $e^{-(\rmg\chi t)^{2}/2\alpha_{\circ}}$ ; and a decoherence time which is inversely proportional to the square root of the mass of field particles.

It will be interesting to generalize this study to the case that the environment is not merely represented by a single-mode bosonic field; and consider some classes of spectral densities for the environment. Also, for the model represented by the Hamiltonian of equation (\ref{1}) at least in principle one should be able to obtain the pointer states of the system and the environment in some other regimes of the parameter space.

%%%%%%%%%%%%%%%%%%%%%%%%%%%%%%%%%%%%%%%%%%%%%%%%%%%%%%%%%%%%%%%%%%%%%%%%%%%%%%%%%%%%%%%%%%%%%%%%%%

\end{document}